\documentclass[aps,prd,twocolumn,showpacs,eqsecnum,nofootinbib]{revtex4}

\newcommand\be{\begin{equation}}
\newcommand\ee{\end{equation}}
\newcommand\ba{\begin{eqnarray}}
\newcommand\ea{\end{eqnarray}}
\newcommand\eq{\begin{equation}}
\newcommand\en{\end{equation}}

\usepackage{graphicx}
\usepackage{dcolumn}
\usepackage{amssymb}

\font\FermiSmallfont=cmssq8 scaled 1200

\def\LANLppthead#1#2#3{
\null
\begin{center}\vskip -1.0truein{\hbox to 7.5truein {
\hfill
\vbox to 1in {\vfill \FermiSmallfont
              \hbox{#1}
              \hbox{#2}
              \hbox{#3}
              \vfill}
}}\vskip-0.0truein\end{center}}

\begin{document}

\LANLppthead {LA-UR-05-4906}{FNAL-PUB-05-259-A}{KAIST-TH/2005-11}

\title{
Parameterizing the Power Spectrum:\\
Beyond the Truncated Taylor Expansion
}

\author{Kevork Abazajian$^1$, Kenji Kadota$^2$
and Ewan D. Stewart$^{3,4}$\footnote{On sabbatical leave from Department of Physics, KAIST, Republic of Korea.} }

\affiliation{$^1$Theoretical Division, MS B285, Los Alamos National Laboratory, Los Alamos, NM~~87545}
\affiliation{$^2$Particle Astrophysics Center, Fermi National Accelerator
Laboratory, Batavia, IL~~60510}
\affiliation{$^3$Department of Physics, KAIST, Daejeon 305-701, Republic of Korea}
\affiliation{$^4$Canadian Institute for Theoretical Astrophysics, University of Toronto, Toronto, ON M5S 3H8, Canada}

\begin{abstract}
  The power spectrum is traditionally parameterized by a truncated Taylor
  series: $\ln \mathcal{P}(k) = \ln \mathcal{P}_* + (n_*-1) \ln(k/k_*) +
  \frac{1}{2} n'_* \ln^2(k/k_*)$. It is reasonable to truncate the Taylor
  series if $|n'_*\ln(k/k_*)| \ll |n_*-1|$, but it is not if $|n'_*\ln(k/k_*)|
  \gtrsim |n_*-1|$. We argue that there is no good theoretical reason to prefer $|n'_*| \ll
|n_*-1|$, and show that current observations are consistent with $|n'_*
\ln(k/k_*)| \sim |n_*-1|$ even for $|\ln(k/k_*)| \sim 1$. Thus, there
are regions of parameter space, which are both theoretically and
observationally relevant, for which the traditional truncated Taylor
series parameterization is inconsistent, and hence it can lead to
incorrect parameter estimations. Motivated by
  this, we propose a simple extension of the traditional parameterization,
  which uses no extra parameters, but that, unlike the traditional approach,
  covers well motivated inflationary spectra with $|n'_*| \sim |n_*-1|$. Our
  parameterization therefore covers not only standard-slow-roll inflation
  models but also a much wider class of inflation models. We use this
  parameterization to perform a likelihood analysis for the cosmological
  parameters.
\end{abstract}
\pacs{98.80.Cq}
\maketitle

\newpage

\section{Introduction}

In estimating the cosmological parameters and the primordial power spectrum
using the observational data, such as WMAP and SDSS
\cite{wmap,Tegmark:2003uf,selj}, one introduces free parameters to
parameterize the power spectrum $\mathcal{P}(k)$.  There is some arbitrariness
in how to do this beyond simply parameterizing the amplitude of a scale
invariant spectrum.  It has become traditional to consider a truncated Taylor
expansion about some particular pivot scale $k_*$, including two derivative
terms, the spectral index $n-1 \equiv d\ln\mathcal{P}/d\ln k$ and its running
$n' \equiv d n/d\ln k$.  This traditional parameterization is motivated by
simplicity and the standard slow-roll approximation.

The standard slow-roll approximation is satisfied by the simplest
single-component models of inflation, but one also has to be prepared for not
so simple single-component models, and it may generically not be satisfied by
multi-component models of inflation.  Currently there is no observational
reason to disfavor multi-component models of inflation, and there are some
theoretical reasons to prefer them in terms of naturalness \cite{suc}.  Thus
using the standard slow-roll approximation to justify a parameterization of the
power spectrum is dangerous.

The truncated Taylor expansion could give a poor approximation if the range of
$k$ of our interest extends far from $k_*$, unless we take a sufficient number
of higher derivative terms in the Taylor expansion.  However, increasing the
number of derivative terms to as many as one desires would not be a practical
approach for parameter estimation.

In this paper, we propose an improved parameterization of the power spectrum,
which is as simple as the traditional truncated Taylor expansion in that it
uses the same number of parameters, reproduces the traditional truncated Taylor
expansion in the standard slow-roll limit, but has a better motivated form
outside of this limit.  Because our parameterization has the same number of
free parameters as the traditional truncated Taylor expansion, it can be
straightforwardly implemented in existing numerical codes.

The layout of our paper is as follows.
The truncated Taylor expansion and its drawbacks are discussed in Sec.~\ref{prob}.
Our proposed parameterization is presented in Sec.~\ref{res1}, and its inflationary motivation in Sec.~\ref{res2}.
In Sec.~\ref{like} cosmological parameter estimation via the Markov-Chain Monte-Carlo (MCMC) method using our parameterization is compared with that using the traditional truncated Taylor expansion.
Conclusion and discussion are in Sec.~\ref{conc}.

\section{The Traditional Approach}
\label{prob}

We describe the traditional approach, explain the weakness of its traditional
justification, and describe some possible undesirable consequences of adopting
it.

\subsection{The traditional approach: a truncated Taylor expansion of the power spectrum}

Traditionally, the deviation from a scale-invariant spectrum is quantified by the tilt $n-1 \equiv d\ln\mathcal{P}/d\ln k$ and its running $n' \equiv d n/d\ln k$ by Taylor expanding in log-log space around a pivot point $\ln k_*$
\begin{equation}
\label{truncated}
\ln \mathcal{P}(k) = \ln \mathcal{P}_* + (n_*-1) \ln \left(\frac{k}{k_*}\right)
+ \frac12 n'_* \ln^2\left(\frac{k}{k_*}\right)
\end{equation}
This traditional truncated Taylor expansion approach assumes the second and higher derivatives of $n$ are negligible, which is not a trivial assumption, especially for $|\ln(k/k_*)| \gtrsim 1$.
Generally, one might expect this to be a good assumption if $|n'_*\ln(k/k_*)| \ll |n_*-1|$, but a bad one if $|n'_*\ln(k/k_*)| \gtrsim |n_*-1|$.
We shall show in Sec.~\ref{like} that current observations are consistent with $|n'_*\ln(k/k_*)| \sim |n_*-1|$, and hence that there is no observational justification for this assumption.

\subsection{The traditional justification of the traditional approach: the standard slow-roll approximation}

The standard slow-roll approximation assumes that the inflationary slow-roll
parameters are both small {\em and\/} slowly varying.  In terms of observable
quantities, smallness of the relevant parameters translates to
\begin{equation}
\label{stand}
|n-1| \ll 1
\end{equation}
which is required by observations, while the slowly varying condition
translates to the hierarchy
\begin{equation}
\label{stand2}
|n-1| \gg |n'| \gg |n''| \gg \ldots
\end{equation}
Thus we see that the validity of the traditional truncated Taylor series
approach is equivalent to the assumption of slowly varying slow-roll
parameters.  Thus if we assume the standard slow-roll approximation, as is
often done, neglection of the higher order derivatives in Eq.~(\ref{truncated})
follows.

In terms of inflationary model building, for example to get enough inflation,
there is no need for the slow-roll parameters to be either small or slowly
varying.  However, as noted above, smallness of the relevant parameters is
required for approximate scale invariance of the spectrum.  In many simple
single-component models of inflation, the requirement of small slow-roll
parameters forces the slow-roll parameters to be slowly varying, but there is
no general requirement of this.  Also, there is no need to restrict to a
single-component inflaton. From the particle theory viewpoint, all scalar
fields are complex in supersymmetry, and there are many scalar fields. From the
inflationary model building point of view, it allows extra freedom to build
more natural models, see for example Ref.~\cite{suc}.  In these multi-component
models, the relevant slow-roll parameters\footnote{In multi-component models,
  there are many slow-roll parameters. The relevant slow-roll parameters are
  the ones that directly affect the spectrum, see Ref.~\cite{new}.}  are small
but, without fine-tuning, most of the irrelevant slow-roll parameters will not
be small. The non-small irrelevant slow-roll parameters then tend to cause the
relevant slow-roll parameters not to vary slowly.  Thus, although standard
slow-roll could be considered the generic, though not exclusive, case in
single-component models, non-slowly varying slow-roll parameters may be the
more generic case in multi-component models.  Thus, if one wants to cover all
reasonable models and hence be able to distinguish amongst them, one should
relax the assumption of slowly varying slow-roll parameters.

The general slow-roll approximation \cite{ewangeneral,new} drops this extra
assumption of slowly varying slow-roll parameters, covering the cases of
\begin{equation}
\label{same}
|n-1| \gtrsim |n'| \gtrsim |n''| \gtrsim \ldots
\end{equation}
Hence the general slow-roll approximation includes Eq.~(\ref{stand2}) as a special case, so that we can test the assumption of the standard slow-roll approximation, rather than assuming it a priori.

\subsection{Possible undesirable consequences of the traditional approach}

If $|n'_*\ln(k/k_*)| \gtrsim |n_*-1|$ then the truncated Taylor series gives a
very unnatural form for the spectrum which is not motivated by any model of
inflation, and which can give misleading parameter estimations.  In general,
the running can become appreciable for high $k$ even if it is negligible for
small $k$ due to the possible $k$ dependence of the running.  Thus, if we
ignore the running of the running, as is done in the traditional truncated
Taylor expansion, the running of the spectrum at high $k$ may be too biased by
the data at low $k$ because the running at high $k$ is forced to be same as
that at low $k$.  In those cases, we should take account of the running of the
running and more generally all the higher order terms.  Considering infinitely
many terms or an infinite number of parameters would, however, be unpractical
in actual parameter estimations, and we shall suggest a more appropriate way of
parameterizing the power spectrum in Sec.~\ref{res}.  Thus taking only the
first few terms in the Taylor expansion could give a poor representation of the
power spectrum and lead us to incorrect parameter estimations.

\section{A Better Approach}
\label{res}

We suggest a different way to parameterize the spectrum, which does not require
any extra parameters compared with the traditional truncated Taylor expansion
parameterization.  In the standard slow-roll limit it reduces to the usual
truncated Taylor series form, but has a more sensible extension beyond the
standard slow-roll limit.

\subsection{Improved parameterization}
\label{res1}

Instead of the truncated Taylor expansion, we parameterize the spectrum as
\begin{equation}
\label{integ}
\ln \mathcal{P}(k) = \ln \mathcal{P}_* + \frac{(n_*-1)^2}{n'_*}  \left[ \left(\frac{k}{k_*}\right)^\frac{n'_*}{n_*-1} - 1 \right]
\end{equation}
and hence the spectral index as
\begin{equation}
\label{good}
n-1 = (n_*-1) \left(\frac{k}{k_*}\right)^\frac{n'_*}{n_*-1}
\end{equation}
and the running as
\begin{equation}
\label{goodprime}
n' = n'_* \left(\frac{k}{k_*}\right)^\frac{n'_*}{n_*-1}
\end{equation}
where, as before, $k_*$ is an arbitrary reference point.  Note that our
parameterization has a simple form and uses the same number of parameters as the
truncated Taylor series of Eq.~(\ref{truncated}).  Expanding Eq.~(\ref{integ})
in the limit $|n'_* \ln(k/k_*)| \ll |n_*-1|$, we see our parameterization
reproduces the standard slow-roll Taylor series
\begin{equation}
\ln \mathcal{P} = \ln \mathcal{P}_* + (n_*-1) \ln\left(\frac{k}{k_*}\right) + \frac{1}{2} n'_* \ln^2\left(\frac{k}{k_*}\right) + \ldots
\end{equation}
of Eq.~(\ref{truncated}), but with a more sensible extension to the domain $|n'_* \ln(k/k_*)| \gtrsim |n_*-1|$.

\subsection{Inflationary motivation of our parameterization}
\label{res2}

The truncated Taylor series and our parameterization are equivalent for $|n'_*\ln(k/k_*)| \ll |n_*-1|$, but our parameterization is also well motivated for $|n'_*\ln(k/k_*)| \gtrsim |n_*-1|$.

Specifically, in the general slow-roll approximation the spectrum for multi-component inflation models is given by \cite{new}
\footnote{Note that in some cases there may be extra terms. See Ref.~\cite{new} for the full story.}
\begin{equation}
\label{mgsr}
\ln\mathcal{P} = \int_0^\infty \frac{d\xi}{\xi} \left[ - k \xi \, W'(k\xi) \right] \left[ \ln\Pi^2 + \frac{2}{3} \frac{\Pi'}{\Pi} \right]
\end{equation}
where $\Pi = \Pi(\ln\xi)$, $\Pi' \equiv d\Pi/d\ln\xi$, and $\xi$ is minus the conformal time: $\xi = - \int dt/a \simeq 1/aH$ where $a$ is the scale factor and $H$ is the Hubble parameter.
The window function $- x \, W'(x)$ is given in Refs.~\cite{ewangeneral,new} and has the properties
\ba
\int^{\infty}_{0} \frac{dx}{x} \left[ - x \, W'(x) \right] = 1
\ea
and
\begin{equation}
\lim_{x \rightarrow 0} \left[ - x \, W'(x) \right] = \mathcal{O}\left(x^2\right)
\end{equation}
$\Pi$ represents the relevant inflationary parameters that directly affect the spectrum and is defined in Ref.~\cite{new}.

At zeroth order, $\Pi^2$ can be regarded as constant and the normalization
property of the window function leads to
\begin{equation}
\ln\mathcal{P} = \ln\Pi^2
\end{equation}

Our parameterization of Eq.~(\ref{integ}) arises from a $\Pi^2$ of the form
\begin{equation}
\ln\Pi^2 = \ln\Pi^2_\infty - B \xi^{-\nu}
\end{equation}
where $\Pi^2_\infty $ is a constant and the second term with the constant coefficient $B$ is assumed to be small, or
\begin{equation}
\label{fpf}
\frac{\Pi'}{\Pi} = \frac{1}{2} \nu B \xi^{-\nu}
\end{equation}
For $\nu < 2$, substituting into Eq.~(\ref{mgsr}) gives
\begin{equation}
\ln\mathcal{P} = \ln\Pi^2_\infty - A k^\nu
\end{equation}
where $A$ is a constant depending linearly on $B$ and non-trivially on $\nu$.
For $\nu \geq 2$, the late time part of the integral dominates and $n-1$ becomes proportional to $k^2$.
See Ref.~\cite{suc} for a more detailed discussion.
Comparing with Eq.~(\ref{integ}), we have
\begin{equation}
\label{forma}
\frac{n'_*}{n_*-1} = \left\{
\begin{array}{ccc}
\nu & \mbox{for} & \nu < 2
\\
2 & \mbox{for} & \nu \geq 2
\end{array}
\right.
\end{equation}
with standard slow-roll corresponding to $\nu \ll 1$. Note that simple single-component inflation models tend to satisfy $|n'|\sim |n-1|^2 \ll |n-1|$. A concrete example of a particle theory motivated inflationary model which gives a spectrum of the form of Eq.~(\ref{forma}) is given in Ref.~\cite{suc}.

Of course, the general slow-roll approximation can also accommodate cases where $\Pi'/\Pi$ cannot be expressed as a power of $\xi$.
In other words, this power law case is still a special case of the more general class of inflation models which the general slow-roll approximation can handle.

\section{Likelihood analysis}
\label{like}

\begin{figure}
\label{aeps}
\centerline{\includegraphics[width=3.35in]{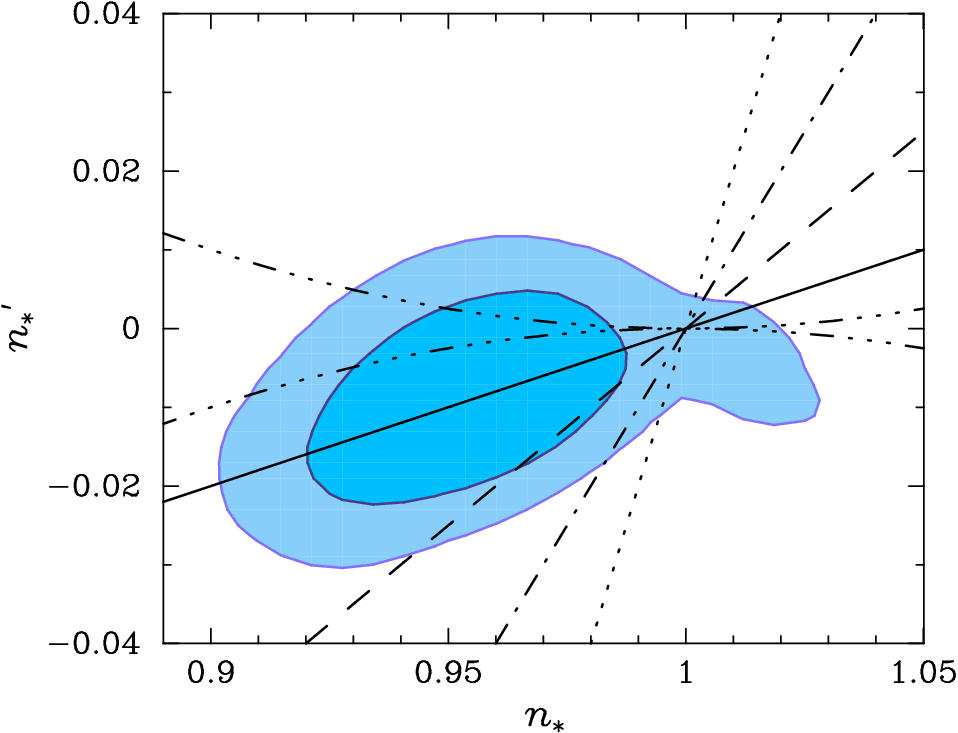}}
\caption{Shown are the likelihood contours of 68.3\% and 95.4\% for ($n_*$,$n'_*$)
  in our parameterization, using observations of the CMB, the SDSS 3D power spectrum
  of galaxies, and the matter power spectrum inferred from the SDSS Lyman-$\alpha$
  forest. Also shown are the lines of $n_*^\prime/(n_*-1) = 0.2, 0.5, 1, 2$ as
  solid, dashed, dot-dashed, and dotted, respectively.  The curves
  $n_*^\prime/(n_*-1)^2 = \pm 1$ are triple-dot dashed.}
\end{figure}

\begin{figure}
\label{numerical}
\centerline{\includegraphics[width=3.35in]{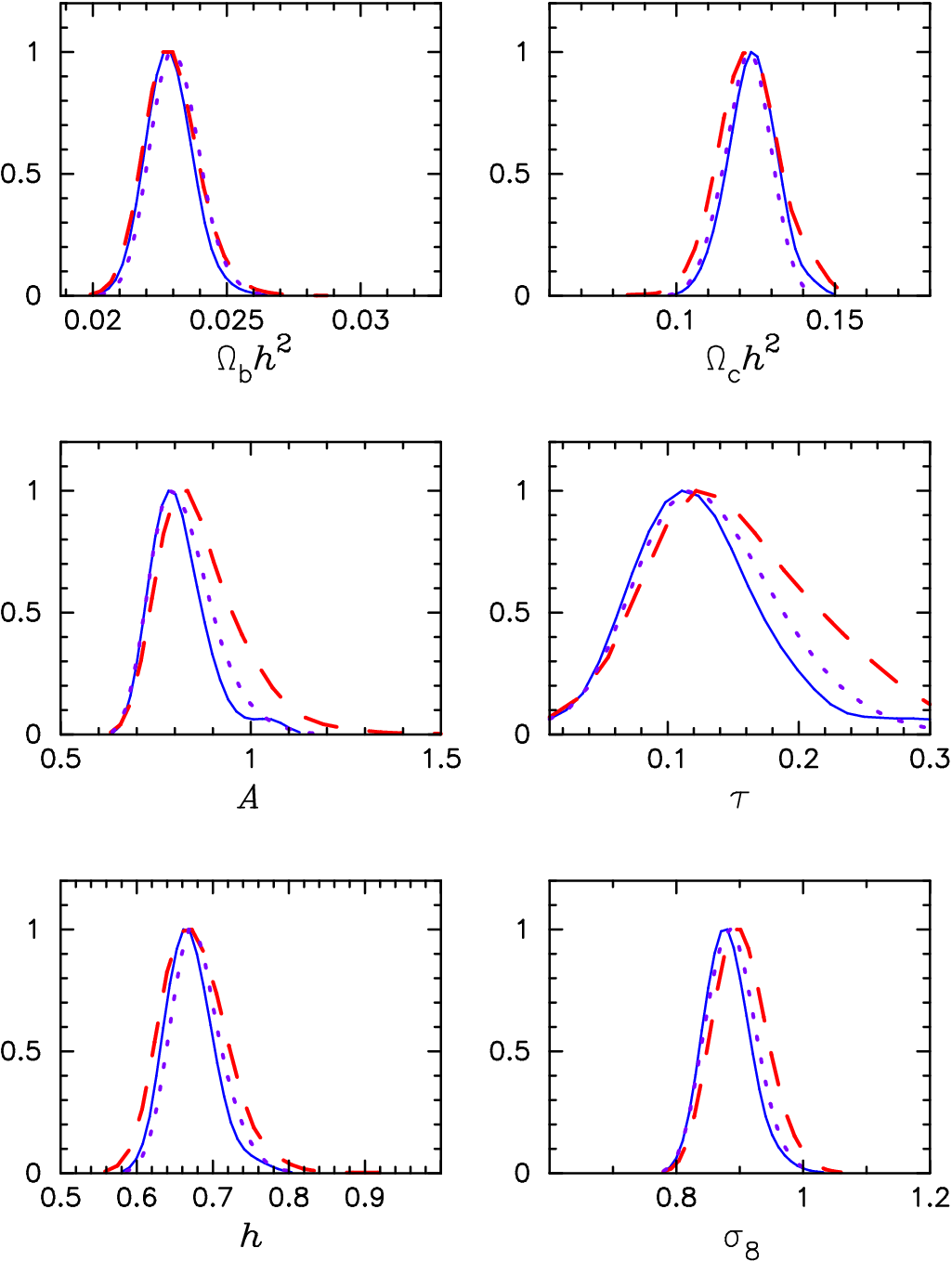}}
\caption{Plotted are the marginalized one-dimensional probability distribution functions
  for cosmological parameter estimation using our parameterization
  (solid), the truncated Taylor expansion (dash), and constant
  spectral index (dotted).}
\end{figure}

\begin{figure}
\label{like_n}
\centerline{\includegraphics[width=3.35in]{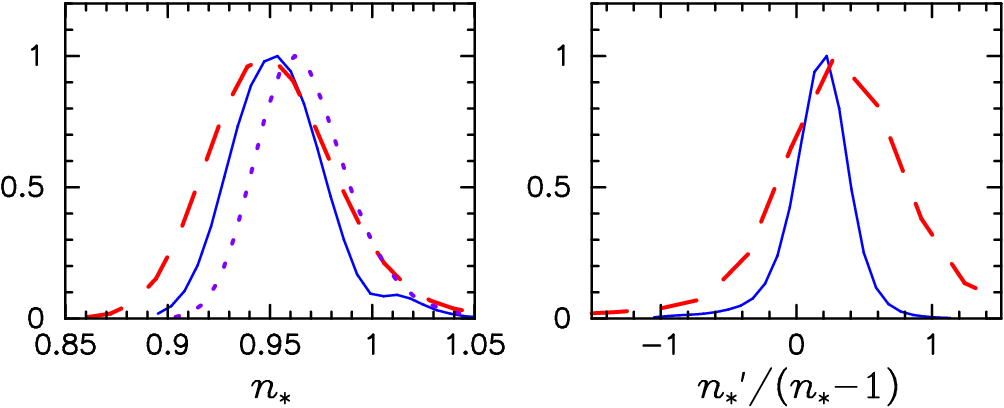}}
\caption{In the left panel, we show the marginalized one-dimensional
  probability distribution functions for $n_*$ for our parameterization (solid),
  for the truncated Taylor expansion (dash), and for constant spectral index
  (dotted). The right panel is that for $n_*^\prime/(n_*-1)$.}
\end{figure}

We perform an estimation of cosmological parameters using different
parameterizations of the primordial perturbation spectrum: (1) our spectral
parameterization given by Eq.~(\ref{good}); (2) the truncated Taylor expansion
given by Eq.~(\ref{truncated}); and (3) the case of constant spectral index.
We model a flat universe with the cosmological parameters $\Omega_b h^2$,
$\Omega_c h^2$, $\Theta_s$, $\ln A$ and $\tau$.  Here, $A$ is related to the
amplitude of curvature perturbations at horizon crossing, $|\Delta_ R|^2 =
2.95\times 10^{-9} A$ at the scale $k_*= 0.05 h\rm /Mpc$.  The angular acoustic peak
scale $\Theta_s$ is the ratio of the sound horizon at last scattering to that
of the angular diameter distance to the surface of last scattering, and is a
useful proxy for the Hubble parameter $H_0 \equiv 100h\rm\ km\ s^{-1}\ Mpc$
\citep{Kosowsky:2002zt}.

We use a Markov-Chain Monte-Carlo (MCMC) technique with a modified form of
CosmoMC \cite{Lewis:2002ah}.  We use the CMB data from first year WMAP
\cite{wmap2}, ACBAR \cite{kuo}, CBI \cite{read} and VSA \cite{dickinson}, the galaxy
power spectrum from SDSS \cite{tmax} and Lyman$-\alpha$ \cite{mcdonald}. Each MCMC analysis
included approximately $10^5$ points, with an acceptance efficiency of
approximately 45\%.  The parameter space was sampled from flat priors
\begin{eqnarray}
0.005\le& \Omega_b h^2 &\le 0.1 \nonumber \\
0.01\le& \Omega_c h^2 &\le 0.99 \nonumber \\
0.005 \le&  \Theta_s &\le 0.1 \nonumber \\
-0.68 \le& \ln A &\le 0.62 \nonumber \\
0.01 \le& \tau &\le 0.3 \ (0.8) \nonumber \\
0.5 \le& n_* &\le 1.5 \nonumber \\
-0.5 \le& n_*' &\le 0.5
\end{eqnarray}
which are all well outside of the regions of high probability, except for the
requirement of $\tau\le 0.3$ for our parameterization case (1), which we include to exclude unphysically high optical depths
otherwise allowable by the data and this form of the power spectrum. We use $\tau\le 0.8$, outside appreciable levels of the PDF for
cases (2) and (3).  Since currently there are only upper limits on the
contribution of tensor perturbations, we do not include tensor
perturbations in our analysis.

The 2-D likelihood of the parameters $n'_*$ and $n_*$ is shown in Fig.~1,
where we also show lines of constant $n_*'/(n_*-1)$
and the curves of $n_*'/(n_*-1)^2 = \pm 1$.
Note that in simple single-component inflation models one often gets $|n'| \sim |n-1|^2 \ll |n-1|$, and our parameterization indicates the current data is consistent with a range well beyond that where this relation holds.
In Fig.~2, we show the cosmological parameter estimation using the three different parameterizations, and in Fig.~3, we show the estimation of $n_*$ and $n_*'/(n_*-1)$.
The central values for the models' parameters, their uncertainty and $\chi^2$ are given in Table~\ref{tab:pars}.

\begin{table}
\caption{\label{tab:pars} Central values, 68.3\% uncertainties, and $\chi^2$
  for the different models of the primordial scalar spectrum.}
\begin{ruledtabular}
\begin{tabular}{lrrr}
 &$n^\prime=0$ & $n^{\prime\prime}=0$ & $n^{\prime\prime} = \frac{{n^\prime}^2}{n-1}$ \\
\hline\\[0.02in]
$\Omega_b h^2$ &
$0.02310^{+0.00087}_{-0.00088}$ &
$0.02298^{+0.00093}_{-0.00094}$ &
$0.02320^{+0.00081}_{-0.00080}$ \\[0.025in]

$\Omega_c h^2$ &
$0.1223^{+0.0070}_{-0.0072}$ &
$0.1233^{+0.0092}_{-0.0092}$ &
$0.1220^{+0.0068}_{-0.0070}$ \\[0.025in]

$\Theta_s$ &
$1.0416^{+0.0041}_{-0.0041}$ &
$1.0420^{+0.0040}_{-0.0040}$ &
$1.0420^{+0.0040}_{-0.0040}$ \\[0.025in]

$\tau$ &
$0.134^{+0.021}_{-0.029}$ &
$0.157^{+0.024}_{-0.038}$ &
$0.138^{+0.024}_{-0.025}$ \\[0.025in]

$A$ &
$0.814^{+0.080}_{-0.029}$ &
$0.854^{+0.104}_{-0.090}$ &
$0.803^{+0.068}_{-0.065}$ \\[0.025in]

$n_*$ &
$0.967^{+0.021}_{-0.022}$ &
$0.953^{+0.029}_{-0.029}$ &
$0.955^{+0.021}_{-0.022}$ \\[0.025in]

$n_*^\prime$ &
$-$ &
$-0.019^{+0.014}_{-0.014}$ &
$-0.0087^{+0.0084}_{-0.0083}$ \\[0.03in]

\hline\\[0.03in]

$\Omega_m$ &
$0.318^{+0.042}_{-0.043}$ &
$0.324^{+0.056}_{-0.055}$ &
$0.330^{+0.044}_{-0.044}$ \\[0.025in]

$h$ &
$0.680^{+0.032}_{-0.032}$ &
$0.678^{+0.039}_{-0.040}$ &
$0.671^{+0.030}_{-0.031}$ \\[0.025in]

$\sigma_8$ &
$0.887^{+0.038}_{-0.038}$ &
$0.902^{+0.041}_{-0.041}$ &
$0.882^{+0.035}_{-0.036}$ \\[0.025in]

$\chi^2$ &
$1664.7$ &
$1663.1$ &
$1664.1$  \\
\end{tabular}
\end{ruledtabular}

\end{table}

In Fig. 1, the 95.4\% CL contour region has a stretched region towards positive
$n_*-1$ and negative $n'_*$ compared with the elliptic shaped 68.3\% CL
contour.  This arises due to the combination of the fit trying to satisfy the
low-$\ell$ multipoles simultaneously with the high-optical depth allowed.

The data used in our analysis covers a $k$ range of $\Delta\ln k \sim 10$.
This means the central values in Table~I giving $|n'_*| \sim 0.2 \, |n_*-1|$ for our parameterization, or $|n'_*| \sim 0.4 \, |n_*-1|$ for the truncated Taylor series, are inconsistent with the range of validity of the truncated Taylor series, as they give $|n'_* \ln(k/k_*)| \sim |n_*-1|$.
Note that the central values of $n'_*$ differ because, for $n'_*/(n_*-1) > 0$, our parameterization alters the power spectrum more strongly for $k > k_*$ than the truncated Taylor expansion.
Despite the lack of robust data at $k$ even larger than the SDSS Lyman-$\alpha$ data, this result is a good illustration of the danger of ignoring the higher derivative terms in the Taylor series when we deal with more precise cosmological data covering a wide range of scales in the future.

\section{Conclusion and Discussion}
\label{conc}

We discussed the possible bias and inconsistency in the cosmological parameter estimation induced by presuming a truncated Taylor series form for the power spectrum.
We proposed an improved form for the power spectrum which is motivated from actual perturbation calculations applicable to a wide class of well motivated inflation models.
The standard slow-roll truncated Taylor series form is just a special limit of our parameterization of the power spectrum.
Our proposed form requires no additional free parameters compared with the traditional truncated Taylor expansion, and can be straightforwardly implemented in existing codes.
Our form for the spectrum is simple
\ba
\label{above1}
\ln \mathcal{P} & = & \ln \mathcal{P}_0 - A k^\nu
\\ \label{above2}
n-1 & = & - \nu A k^\nu
\ea
with only three constant free parameters, $\mathcal{P}_0$, $A$ and $\nu$.
There exist several possible ways to parameterize this form.
The parameterization used in this paper, Eq.~(\ref{integ}), was motivated by clarity in the comparison with the truncated Taylor expansion.
The standard slow-roll cases correspond to $\nu = n'_*/(n_*-1) \ll 1$.
In performing the likelihood analysis via MCMC, we found the central values for $n'_*/(n_*-1)$ were inconsistent with $|n'_* \ln(k/k_*)| \ll |n_*-1|$, the basic assumption behind the truncated Taylor series.

Further to the technical discussion on the form of the power spectrum using the general slow-roll formula in Sec.~\ref{res2}, let us point out that the form we presented in this paper should be regarded as an asymptotic form for small $A k^\nu$.
This asymptotic form is sufficient for the range of our analysis in this paper, \mbox{i.e.} the data up to $k\sim$ 6h/Mpc from Lyman$-\alpha$.
If we can include data at higher $k$, well beyond the Lyman-$\alpha$ range, covering a change in $\ln\mathcal{P}$ of order unity, we may need to use the original form of Ref.~\cite{suc} without asymptotic approximation.
Otherwise $\mathcal{P}$ would decrease too rapidly for higher $k$ (assuming $\nu > 0$).
Alternatively we may soften the form of our parameterization, for example as
\ba
\ln\mathcal{P} & = & \ln\mathcal{P}_0 - \ln \left( 1 + A k^\nu \right)
\\
n-1 & = & - \frac{A \nu k^\nu}{1 + A k^\nu}
\ea
so that it gives Eqs.~(\ref{above1}) and~(\ref{above2}) for small $A k^\nu$, and
\ba
\ln\mathcal{P} & = & \ln\mathcal{P}_* - \nu \ln\left(\frac{k}{k_*}\right)
\\
n-1 & = & - \nu
\ea
for large $A k^\nu$, or ideally add more parameters.

Simple single-component inflation models require the standard slow-roll approximation to produce a flat spectrum, but this is not the case for multi-component inflation models.
In this sense, our parameterization would be of great interest for multi-component inflation models, such as given in Ref.~\cite{suc}.

Our parameterization, the traditional truncated Taylor expansion, and the case of constant spectral index led to quite similar, though not identical, results for the cosmological parameter estimations using the currently available data dominated by large scale observations such as CMB and galaxy surveys.
This indicates that currently the running is not crucial to the cosmological parameter estimations, but this may change with reduced error bars in the near future \cite{leach2}.
Application of a parameterization such as ours would also be important for modeling very small scale structure, such as the first objects in universe \cite{first} which could extend the range up to $k\sim 10^6$ h/Mpc.

\subsection*{Acknowledgments}
We thank Scott Dodelson, Zoltan Haiman, Wayne Hu, Lam Hui, Pat McDonald and
Jochen Weller for useful discussions. KK thanks Jochen Weller for the early
stage of collaboration. KA is supported by Los Alamos National Laboratory
(under DOE contract W-7405-ENG-36).  KK is supported by Fermilab (under DOE
contract DE-AC02-76CH03000) and by NASA grant NAG5-10842.  EDS is supported by
ARCSEC funded by the Korea Science and Engineering Foundation and the Korean
Ministry of Science, Korea Research Foundation grant KRF PBRG 2002-070-C00022,
and Brain Korea 21.


\end{document}